\documentclass[12pt,letter]{revtex4}
\usepackage{setspace}

\begin{document}

\title{\bf Quantum Theory:  Exact or Approximate?}
\author{Stephen L. Adler} \email{adler@ias.edu}
\affiliation{Institute for Advanced Study, Einstein Drive,
Princeton, NJ 08540, USA}

\author{Angelo Bassi}
\email{bassi@ts.infn.it} \affiliation{Department of Theoretical
Physics, University of Trieste, Strada Costiera 11, 34014 Trieste,
Italy} \affiliation{Istituto Nazionale di Fisica Nucleare, Trieste
Section, Via Valerio 2, 34127 Trieste, Italy}

\maketitle

Quantum mechanics has enjoyed a multitude of successes since its
formulation in the early twentieth century. It has explained the
structure and interactions of atoms, nuclei, and subnuclear
particles, and has given rise to revolutionary new technologies.  At
the same time, it has generated puzzles that persist to this day.

These puzzles are largely connected with the role that measurements
play in quantum mechanics~\cite{L}.  According to the standard
quantum postulates, given the Hamiltonian, the wave function of
quantum system evolves by Schr\"odinger's equation in a predictable,
deterministic way. However, when a physical quantity, say z-axis
spin, is ``measured'', the outcome is not predictable in advance. If
the wave function contains a superposition of components, such as
spin up and spin down, which each have a definite spin value,
weighted by coefficients $c_{\rm up}$ and $c_{\rm down}$, then a
probabilistic distribution of outcomes is found in repeated
experimental runs. Each repetition gives a definite outcome, {\it
either} spin up or spin down, with the outcome probabilities given
by the absolute value squared of the corresponding coefficient  in
the initial wave function. This recipe is the famous Born rule.  The
puzzles posed by quantum theory are how to reconcile this
probabilistic distribution of outcomes with the deterministic form
of Schr\"odinger's equation, and to understand precisely what
constitutes a ``measurement''.  At what point do superpositions
break down, and definite outcomes appear? Is there a quantitative
criterion, such as size of the measuring apparatus, governing the
transition from coherent superpositions to definite outcomes?

These puzzles have inspired a large literature in physics and
philosophy. There are two distinct approaches.  One is to assume
that quantum theory is exact, but that the interpretive postulates
need modification, to eliminate apparent contradictions.  Many
worlds, decoherent histories, Bohmian mechanics, and quantum theory
as information, all  fall in this category.  Although their
underlying mathematical formulations differ, empirically they are
indistinguishable, since they  predict the same experimental results
as does standard quantum theory.

The second approach is to assume that quantum mechanics is not
exact, but instead is a very accurate approximation to a deeper
level theory, which reconciles the deterministic and probabilistic
aspects.  This may seem radical, even heretical, but looking back in
the history of physics, there are precedents.  Newtonian mechanics
was considered to be exact for several centuries, before being
supplanted by relativity and quantum theory, to which classical
physics is an approximation. But apart from this history, there is
another important motivation for considering modifications of
quantum theory. This is to give a quantitative meaning to
experiments testing quantum theory, by having an alternative theory,
making  predictions that differ from those of standard quantum
theory, to which these experiments can be compared.

Although a modification of quantum theory may ultimately require a
new dynamics, we  focus here on phenomenological approaches, that
look for modifications of the Schr\"odinger equation that describe
what happens in measurements.  A successful phenomenology must
accomplish many things:  (1) It must explain why repetitions of the
same measurement lead to definite, but differing, outcomes.  (2) It
must explain why the probability distribution of outcomes is given
by the Born rule.  (3) It must permit quantum coherence to be
maintained for atomic and  mesoscopic systems, while predicting
definite outcomes for measurements with realistic apparatus sizes in
realistic measurement times. (4)  It should conserve overall
probability, so that particles do not spontaneously disappear. (5)
It should not allow superluminal transmission of signals, while
incorporating quantum nonlocality.

It is not obvious that a phenomenology should exist that satisfies
these requirements, but remarkably, through work over the last two
decades, one does.  One ingredient is the observation that rare
modifications, or ``hits'', localizing the wave function, will not
alter atomic-level coherences, but when accumulated over a
macroscopic apparatus can lead to definite outcomes which differ
from run to run~\cite{GRW}. A second ingredient is  the observation
that the classic ``gambler's ruin'' problem in probability theory
gives a mechanism that can explain the Born rule governing outcome
probabilities~\cite{P}, as follows.  Suppose that Alice and Bob each
have a stock of pennies, and flip a fair coin.  If the coin shows
heads, Alice gives Bob a penny, while if the coin shows tails, Bob
gives Alice a penny.  The game ends when one player has all the
pennies and the other has none.  Mathematical analysis shows that
the probability of each player winning is proportional to their
initial stake of pennies.  Map the initial stake into the modulus
squared of the initial spin component coefficient, and one has a
mechanism for obtaining the Born rule.

The combination of these two ideas leads to a definite model, called
the Continuous Spontaneous Localization (CSL) model~\cite{GPR}, in
which a Brownian motion noise term coupled to the local mass density
is added to the Schr\"odinger equation, with a nonlinear noise
squared term included to  preserve wave function normalization.  The
standard form of this model has a linear evolution equation for the
noise averaged density matrix, forbidding superluminal
communication.  Other versions of the model exist, as reviewed
in~\cite{BG,P'}, and a  pre-quantum dynamics has been proposed for
which this model would be a natural phenomenology~\cite{A}.

The CSL model has two intrinsic parameters.  One is a rate parameter
$\lambda$, with dimensions of inverse time, governing the noise
strength.  The other is a length $r_C$, which can be interpreted as
the spatial correlation length of the noise field. Conventionally,
$r_C$ is taken as $10^{-5} {\rm cm}$, but any length within an order
of magnitude of this would do.  Demanding that a pointer composed of
$\sim 10^{15}$ nucleons should settle to a definite outcome in $\sim
10^{-7}$ seconds or less, with the conventional $r_C$, requires that
$\lambda$ should be greater than  $\sim  10^{-17} {\rm s}^{-1}$.
That is, requiring that measurements happen in reasonable times with
a minimal apparatus places a {\it lower bound} on $\lambda$.   If
one requires  that latent image formation in photography, rather
than subsequent development, constitutes a measurement, the fact
that few atoms move significant distances in latent image formation
requires an enhanced  lower bound for $\lambda$ a factor of $\sim
10^8$ larger~\cite{A'}.  Note that the Hubble constant is  $\simeq 2
\times 10^{-18} {\rm s}^{-1}$, so the conventional value of
$\lambda$ could be compatible with a cosmological origin of the
noise field, which seems unlikely if $\lambda$ were much enhanced.

An {\it upper bound} on $\lambda$ is placed by the requirement that
apparent violations of energy conservation, taking the form of
spontaneous heating produced by the noise, should not be too large;
the best bound  comes from heating of the intergalactic
medium~\cite{A'}. Spontaneous radiation from atoms places another
stringent bound~\cite{F}, which can however be evaded if the noise
is non-white, with a frequency cutoff~\cite{AR,AB}.  Laboratory  and
cosmological bounds on $\lambda$ (for $r_C = 10^{-5} {\rm cm}$) are
summarized in the Table.

Accurate tests of quantum mechanics that have been performed or
proposed  include diffraction of large molecules in fine mesh
gratings~\cite{AZ}, and a  cantilever mirror incorporated into an
interferometer~\cite{M}.  The Table shows the current limit on
$\lambda$ that has been obtained to date in fullerene diffraction,
and the limit that would be obtained if the proposed cantilever
experiment attains full sensitivity~\cite{BIA}.  To confront the
conventional (enhanced) value of $\lambda$, one would have to
diffract molecules a factor of $10^6$ ($10^2$) larger than
fullerenes.

In terms of distinguishing between conventional quantum theory, and modified
quantum theory as given by the CSL model, experiments do not yet tell us
whether quantum theory is exact, or approximate. Future lines of research
include refining the sensitivity of current experiments, to reach the
capability of making this decision, and achieving a deeper understanding of the
origin of the CSL noise field.

\vskip 1.5cm

\setstretch{0.9}
\begin{center}
\begin{table}
\begin{tabular}{|c|c||c|c|} \hline
\multicolumn{4}{c}{\bf } \\
\multicolumn{4}{c}{\bf Upper bounds on the parameter $\lambda$ of the CSL model} \\
\multicolumn{4}{c}{\bf (with noise correlation length $r_C \sim 10^{-5}$ cm)} \\
\multicolumn{4}{c}{\bf } \\
\hline \hline
& {\footnotesize\bf Distance (in orders of} & & {\footnotesize\bf
Distance (in orders of} \\
$\qquad${\footnotesize\bf Laboratory}$\qquad$ & {\footnotesize\bf
magnitude) from the} & $\qquad${\footnotesize\bf
Cosmological}$\qquad$ & {\footnotesize\bf magnitude) from
the} \\
{\footnotesize\bf Experiments} & {\footnotesize\bf conventional
value} & {\footnotesize\bf Data} & {\footnotesize\bf
conventional value} \\
& {\footnotesize $\lambda \sim  10^{-17} \rm{s}^{-1}$} &
& {\footnotesize $\lambda \sim  10^{-17} \rm{s}^{-1}$} \\
\hline
{\footnotesize Fullerene} & & {\footnotesize Dissociation} &
\\
{\footnotesize  diffraction} & 13 & {\footnotesize of cosmic}
& 17 \\
{\footnotesize experiments} & & {\footnotesize hydrogen} & \\
\hline
{\footnotesize Decay of} & & {\footnotesize Heating of } &  \\
{\footnotesize supercurrents} & 14 & {\footnotesize  intergalactic medium} & 8  \\
{\footnotesize (SQUIDS)} & & {\footnotesize (IGM)} & \\ \hline
{\footnotesize Spontaneous} & & {\footnotesize Heating of } & \\
{\footnotesize X-ray emission} & 6 & {\footnotesize interstellar  dust} & 15 \\
{\footnotesize from Ge} & & {\footnotesize grains} & \\ \hline
{\footnotesize Proton} & & \multicolumn{2}{|c}{} \\
{\footnotesize decay} & 18 & \multicolumn{2}{|c}{} \\
& & \multicolumn{2}{|c}{} \\ \cline{1-2}
{\footnotesize Mirror cantilever} & & \multicolumn{2}{|c}{} \\
{\footnotesize interferometric} & 9 & \multicolumn{2}{|c}{} \\
{\footnotesize experiment} & & \multicolumn{2}{|c}{} \\
\cline{1-2}
\end{tabular}
\vskip 1.0cm \caption{The table gives upper bounds on $\lambda$ from
laboratory experiments and cosmological data, compared with the
conventional CSL model  value $\lambda \sim  10^{-17} {\rm s}^{-1}$.
Reducing the numbers by 8 gives the distance of each bound from the
enhanced value $\lambda  \sim 10^{-9} {\rm s}^{-1}$ obtained if one
assumes that latent image formation constitutes measurement. The
X-ray emission bound excludes an enhanced $\lambda$ for white noise,
but this constraint is relaxed if the noise spectrum is cut off
below $10^{18} {\rm s}^{-1}$. Large molecule diffraction would
confront the CSL value of $\lambda$ for molecules heavier  than
$\sim 10^9$ Daltons, and would confront the enhanced $\lambda$ for
molecular weights greater than $\sim 10^5$ Daltons. (The molecular
diffraction bound on $\lambda$ decreases as the inverse square of
the molecular weight, provided the molecular radius is less than
$r_C$.)}
\end{table}
\end{center}

\end{document}